\documentclass[final,5p,times,twocolumn,authoryear]{elsarticle}

\usepackage{amssymb}
\usepackage{lipsum}
\usepackage{hyperref}
\usepackage{listings}
\usepackage{xcolor}
\usepackage{ragged2e}
\usepackage{array}
\usepackage{pdflscape}
\usepackage{comment}

\journal{arXiv}

\begin{document}

\begin{frontmatter}

\title{Co-produced decentralised surveys as a trustworthy vector to put employees' well-being at the core of companies' performance}

\author[first]{Ad\`ele Br\'eart De Boisanger}
\author[first]{Wendy Sims-Schouten}
\author[first,second]{Francois Sicard \corref{cor1}}
\cortext[cor1]{Corresponding author}
\ead{f.sicard@ucl.ac.uk}
\affiliation[first]{organisation={Department of Arts and Sciences},
            addressline={University College London}, 
            city={London},
            postcode={WC1E 7JG}, 
            country={UK}}
\affiliation[second]{organisation={Centre for Blockchain Technologies},
            addressline={University College London}, 
            city={London},
            postcode={WC1E 6BT}, 
            country={UK}}

\begin{abstract}

Assessing employees' well-being has become central to fostering an 
environment where employees can thrive and contribute to companies' 
adaptability and competitiveness in the market. 
Traditional methods for assessing well-being often face significant challenges, with a major issue being the lack of trust and confidence employees may have in these processes. Employees may hesitate to provide honest feedback due to concerns not only about data integrity and confidentiality, but also about power imbalances among stakeholders. 
In this context, blockchain-based decentralised surveys, leveraging the immutability, transparency, and pseudo-anonymity of blockchain technology, offer significant improvements in aligning responsive actions with employees' feedback securely and transparently. Nevertheless, their implementation raises complex issues regarding the balance between trust and confidence.  
While blockchain can function as a confidence machine for data 
processing and management, it does not inherently address the equally important cultural element of trust. To effectively integrate blockchain technology into 
well-being assessments, decentralised well-being surveys must be supported by cultural practices that build 
and sustain trust. Drawing on blockchain technology management and relational cultural theory, we explain how trust-building can be achieved through the co-production of 
decentralised well-being surveys, which helps address power imbalances between the implementation team and stakeholders. Our goal is to provide a dual cultural-technological framework along with conceptual clarity on how the technological implementation 
of confidence can connect with the cultural development of trust, 
ensuring that blockchain-based decentralised well-being surveys are not only secure and reliable but also perceived as trustworthy vector to improve workplace conditions.
\end{abstract}

\begin{keyword}

blockchain technologies \sep co-production \sep relational cultural theory \sep employees well-being \sep organisational performance 



\end{keyword}

\end{frontmatter}

\section{Introduction}
\label{introduction}

The concept of workplace well-being has undergone a significant evolution from its 19th century roots, which primarily addressed the reduction of physical risks, to a modern framework that also considers the psychological aspects of work life \cite{Hammer2021}. Agencies such as the U.S. Occupational Safety and Health Administration have expanded their regulatory scope to include psychosocial risks alongside traditional physical hazards \cite{Neilan2020}. Likewise, the International Labour organisation now interprets workplace well-being comprehensively, recognizing its connection to the entirety of working life, from the physical working conditions to the emotional states of employees regarding their jobs \cite{Foncubierta-Rodríguez2024}. 
Historically, organisations treated employee well-being as an optional extra, leaving it largely the responsibility of individual employees. However, the turn of the millennium marked a significant shift, with growing expectations for company leadership to proactively manage and implement well-being initiatives due to the emerging connection between employee well-being and organisational performance \cite{Macvicar2022}. This shift toward organisational accountability has become even more pronounced with the growing expectations of Generation Z, a demographic that places unprecedented value on emotional balance and mental health in the workplace \cite{Hilton2024}. These expectations have been further accelerated by the challenges faced by the global workforce during the COVID-19 pandemic \cite{Macvicar2022}.

The idea that employees' well-being is a predictor of job performance has been supported by Wright and coworkers since the early 1990s \cite{Wright1993, Cropanzano1999}. Employees struggling with poor mental health may engage in antagonistic behavior toward their peers \cite{Bolger1991}, leading to workplace conflict and a breakdown in essential teamwork. On the other hand, according to Fredrickson and Branigan's \textit{Broaden-and-Build} theory \cite{Fredrickson2005}, positive workplace emotions can enhance cognitive function and social skills, ultimately fostering personal growth and improving social interactions \cite{Barsade1993}. This positive dynamic can help mitigate the significant financial and human costs associated with mental health issues in the workplace \cite{George1992}.
These studies support the idea that organisations prioritising employee comprehensive well-being can achieve lower staff turnover and burnout rates, alongside higher productivity and engagement levels \cite{Fabius2016, Grossmeier2016}. Consequently, there has been a strategic shift towards viewing employee well-being as an essential, rather than optional, aspect of business strategy. Leaders, such as corporate executives and human resources managers, who have a significant impact on business decisions, are increasingly committed to comprehensive well-being strategies that are proactive and organisation-wide \cite{Macvicar2022}. For example, recent studies using simulations and historical market performance data show that companies with robust employee well-being programs tend to outperform others in the market. A notable example is a group of 45 publicly traded companies with high health and wellness scores (HERO Scorecard), which showed a 235\% increase in value, exceeding the S\&P 500's 159\% increase over the same six-year period \cite{Grossmeier2016}.\\

Not only is employee well-being important for company performance, but it is also progressively becoming a pivotal factor at a competitive funding level, particularly concerning environmental, social, and governance (ESG) criteria. For companies, ESG criteria are no longer ancillary; they are central to securing investment, fostering consumer trust, and achieving competitive differentiation \cite{Pineau2022}. Within the tapestry of ESG, the 'Social' component is undergoing a significant evolution, with employee well-being poised to become a more pronounced criterion, primarily due to its strong link with performance. As comprehensive well-being — defined as a holistic approach that considers physical, mental, emotional, social, and financial dimensions — rises to the forefront of global social issues, especially post-pandemic, ESG reporting is beginning to reflect these nuances. Investors are increasingly directing capital toward companies that demonstrate responsible stewardship of environmental resources, uphold social equity, and practice transparent and ethical governance \cite{Pineau2022}. 
In this context, comprehensive well-being plays a crucial role in addressing the underlying issues associated with stigmas in the workplace, such as race-based discrimination or gender equity, and is not just social issues; it has also become part of governance challenges that can alter a company's perception and performance \cite{Macvicar2022}. 

With the social component of ESG rapidly gaining momentum, one of the biggest challenges for businesses has become accurately measuring and reporting it, including metrics and approaches that reflect comprehensive workplace well-being, which increasingly requires robust data collection systems supported by digital technologies. This ESG digitalisation involves leveraging digital tools, such as web-based platforms and mobile-based applications, supported by data analytics. 
However, relying on these digital tools for well-being assessment poses significant challenges due to the private nature of comprehensive well-being and the general reluctance among employees to openly disclose sensitive personal information, specifically those related to mental and emotional aspects of well-being often heavily burdened by stigmas.
This is compounded by the opacity and the prevalent lack of trust in the systems used to evaluate employee well-being, which often fail to provide the transparency, anonymity, and security required to encourage employees to come forward, such as lack of consideration regarding whether their concerns will be taken seriously, or fear of identity exposure and potential retaliation, that can significantly exacerbate stigmas, and can prevent honest communication and feedback \cite{JeanKing2021}. \\

To address these limitations, the recent advent of blockchain technology is set to offer a promising shift toward more secure, transparent, and anonymous methods of managing well-being assessments \cite{SHINE2021, Rahman2022}. This technology, often hailed as a confidence machine, enhances the reliability of, and introduces new dynamics to data management processes \cite{DeFilippi2020}. While decentralised well-being surveys are slowly trying to establish themselves as legitimate alternatives to centralised digital approaches, they primarily rely on traditional well-being surveys whose design framework does not genuinely integrate with blockchain technology. In particular, while blockchain can function as a confidence machine for reliable data processing, it does not inherently address the equally important cultural aspect of trust needed by employees to express genuine concerns and needs. This oversight creates both technological and cultural disparities that can render this new approach ineffective, as employees may remain hesitant to engage openly without the assurance that their stigmas will be appropriately considered and that power imbalances between stakeholders will be effectively addressed.

Here, we discuss why decentralised well-being surveys must be supported by cultural practices that build and sustain trust in order to effectively integrate blockchain technology into well-being assessments. Drawing on relational cultural theory, we explain how trust-building can be culturally achieved through the co-production of decentralised well-being surveys, which helps address power imbalances between the implementation team and stakeholders. 
Our goal is to provide a dual cultural-technological framework along with conceptual clarity on how the technological implementation of confidence can connect with the cultural development of trust, ensuring that blockchain-based decentralised well-being surveys are not only secure and reliable, but also perceived as trustworthy tools for improving workplace conditions.

\section{Well-being assessment and the issue of trust and confidence in the digital realm}

Traditional approaches to assessing well-being have predominantly utilised self-report measures, aiming to capture individuals' subjective evaluations of their own quality of life \cite{Ruggeri2020}. However, the endeavour to quantify such a complex and inherently subjective construct has inevitably led to the proliferation of a wide array of instruments and methodologies, each with its own theoretical underpinnings and focus areas \cite{Linton2016}. This diversity reflects the rich and nuanced nature of well-being itself, which encompasses not only hedonic aspects, such as happiness and pleasure, but also eudaimonic dimensions, concerning psychological and existential fulfillment \cite{Ruggeri2020}. Despite the continuous development and refinement of these instruments, no single tool has gained universal acceptance, a testament to the ongoing debate over what constitutes well-being and how best to measure it \cite{Ruggeri2020}. 

\subsection{Well-being assessment in the digital realm}
The many quantitative and qualitative methods for well-being assessment are designed to gauge the multifaceted aspects of employee well-being, ranging from job satisfaction and emotional health to physical well-being and work-life balance \cite{Linton2016}. These are crucial for understanding the factors that contribute to a productive, engaged, and healthy workforce \cite{Zheng2015}. For example, interviews and focus group discussions offer qualitative insights into employee well-being. Through open-ended questions, employees can discuss their experiences, challenges, and suggestions for workplace improvements. On the other hand, regular feedback mechanisms, such as suggestion boxes or employee forums, which can be facilitated through online platforms, allow employees to provide continuous input regarding their well-being and workplace conditions. 
However, the ongoing digitalisation and automation of workplace assessments increasingly prioritise the use of quantitative  methods for evaluating employees' well-being, which aligns with the growing emphasis on metrics in the digital fabric of modern societies \cite{Kryzhanovskij2021}. Such approaches present opportunities for integration into modern digital frameworks, such as blockchain, which can leverage automated tools like smart contracts, i.e. self-executing pieces of code deployed on the blockchain that enforce predefined conditions. This becomes increasingly relevant as organisational structures evolve in tandem with technological advancements.

The most widely used quantitative tools for well-being assessment are standardised surveys and questionnaires, which can be customized to measure various dimensions of well-being  
\cite{Linton2016}. 
The effectiveness of these tools, developed through extensive theoretical and empirical work, lies in their ability to efficiently and automatically gather extensive data from a broad segment of the workforce, providing valuable insights into job satisfaction, workplace environment, mental health, and overall well-being \cite{Mills2005}. 
Well-being assessment methods based on numerical scale responses (e.g., Likert scales), which are primarily used or can be adapted for the workplace, include among others:
\begin{itemize}
    \item Job Satisfaction Survey (JSS): measures employee satisfaction across job-related factors such as pay, promotion, and supervision \cite{Spector1985}.
    \item Psychological Well-being Scale (PWB): focuses on eudaimonic well-being, assessing personal growth, autonomy, and life purpose \cite{Ryff1989}.
    \item Work-Related Quality of Life scale (WRQoL): Evaluates quality of life related to work, including job satisfaction, stress, and work-life balance \cite{VanLaar2007}.
    \item Warwick-Edinburgh Mental Well-being scale (WEMWBS): Assesses overall mental well-being, focusing on positive aspects of mental health \cite{Tennant2007}.
    \item Worplace PERMA Profiler: Uses the PERMA model to assess positive emotion, engagement, relationships, meaning, and accomplishment in the workplace \cite{Butler2016}.
    \item Occupational Stress index (OSI): Evaluates the level of occupational stress and its impact on employee health \cite{Belki2008}.
    \item Copenhagen Psychosocial Questionnaire (COPSOQ): Measures psychosocial risks in the workplace, including job demands, control, and social support \cite{Kristensen2005}.
    \item Maslach Burnout Inventory (MBI): Assesses burnout by measuring emotional exhaustion, depersonalisation, and reduced personal accomplishment \cite{Maslach1997}.
    \item General Health Questionnaire (GHQ): Screens for general psychological health and potential mental health issues in employees \cite{Goldberg1979}.
    \item SF-36 Health Survey: Measures health-related quality of life across both physical and mental health dimensions \cite{Brook1979}.
    \item Job Demands-Resources Model Survey(JD-R): Assesses the balance between job demands and resources to predict well-being and burnout \cite{Schaufeli2014}.
    \item Survey of Perceived organisational support: (SPOS): Measures employees’ perceptions of how much the organisation values their contributions and well-being \cite{Worley2009}.
    \item Harvard Well-Being Assessment: A comprehensive tool assessing well-being across multiple domains like happiness, health, meaning, and relationships \cite{WeziakBialowolska2021}.
\end{itemize}

The availability of such comprehensive methods enables organisations to assess well-being pragmatically, guiding the development of evidence-based interventions aimed at fostering a healthier, more engaged workforce. 
Notably, the effectiveness of these questionnaires is underscored by the use of rigorous psychometric models rooted in extensive academic research, and by predefined categories such as job satisfaction, psychological well-being, and employee engagement, ensuring applicability across diverse organisational and cultural contexts \cite{Ruggeri2020}. However, their design often overlooks the reflective involvement of different stakeholders, as can be achieved, for example, through co-production \cite{Ostrom1996}. As a result, they may not adequately account for the diverse stigmas and unique lived-experiences of employees, potentially failing to address power imbalances and the specific contextual factors that influence well-being in the workplace.

\subsection{Importance of anonymity and confidentiality in the digital realm}

The digital implementation of well-being surveys and questionnaires in the workplace often encounters a significant hurdle that can impede the effectiveness and accuracy of these assessments, namely the concerns surrounding anonymity and confidentiality \cite{Fisher2020}. While the questionnaires themselves are designed to capture various facets of employee well-being, the manner in which they are administered, and the ensuing handling of data are critical in ensuring that the responses obtained are both transparent and representative of the true state of employee’s wellbeing \cite{Kaiser2009}.
The majority of the time, companies delegate the task to external private institutes, which specialise in collecting and processing the data garnered from surveys and questionnaires \cite{Iphofen2011}. External entities are often preferred by employees who may feel more at ease disclosing information to independent bodies rather than internal teams. However, the challenge remains to ensure that the aggregation and handling of sensitive data occur in a manner that upholds confidentiality and anonymity, 
regardless of whether these processes are managed by private organisations or academic institutions \cite{Fisher2020}. Therefore, while these external entities often possess the necessary technological tools and knowledge to administer or even design sophisticated surveys, this arrangement introduces unique challenges that can impact the trust in the organisation and the efficacy of the well-being assessment process. 

When companies outsource well-being assessments to private institutes, a primary concern is whether all collected data are accurately processed and considered in the analysis \cite{Iphofen2011}. The fear that certain responses might be inadvertently or voluntarily omitted or lost during data transfer or processing can lead companies to question the completeness of the insights derived from these assessments. This concern often stems from a lack of direct oversight over the data processing methods employed by external vendors. From the employee perspective, scepticism may arise regarding how their feedback is handled by an external entity and whether their individual voices will truly be heard and acted upon \cite{Kaiser2009}. The detachment from the internal processes of their organisation can lead to doubts about the impact of their contributions, diminishing their sense of agency and the perceived value of participating in these assessments \cite{Fisher2020}. 

In this perspective, the assurance of anonymity and confidentiality stands as a cornerstone in creating an environment conducive to trust and open communication within organisations. This foundation of trust is essential for encouraging employees to share their honest feedback on well-being without the looming fear of potential identification and the repercussions that might follow. Because the surveys are conducted by the company itself, the fear of being singled out for candid feedback can create an atmosphere where employees prefer to withhold their true feelings or to present a facade that aligns with what they perceive as organisationally acceptable \cite{Rahman2022}. In situations where anonymity and confidentiality are not convincingly assured, employees are naturally inclined towards self-censorship. This act of self-preservation not only dilutes the authenticity of the feedback but also undermines the organisation's ability to grasp the true state of employee well-being and therefore collect significant data. The accuracy of well-being assessments is fundamentally dependent on the genuine and uninhibited participation of employees \cite{Leimanis2021}. 

\subsection{Revising the role of Trust and confidence in Digital well-bing assessment}
In this context, revisiting the fundamentals of psychology through the concepts of trust and confidence can be instrumental in illuminating the gaps present in current methods of assessment and the evaluation of a new technology. The nuanced debate between trust and confidence reveals intricate dimensions of human interaction, especially in contexts characterized by uncertainty and dependence on others for beneficial outcomes.
Trust, as elucidated by Gambetta \cite{Gambetta2000}, is a complex interplay between expectations and vulnerabilities, where the trustor makes a conscious decision to rely on another entity under conditions of uncertainty. This decision to trust is not taken lightly; it embodies a calculated assessment of potential actions and outcomes. However, this choice carries inherent risks, making the trustor vulnerable to the possibility of betrayal or disappointment \cite{Smith2005}. When the trustor opts for trust, they implicitly assume responsibility for this decision, understanding that should their judgment prove misguided, they will bear a portion of the blame for any negative consequences \cite{Smith2005}. 

Unlike confidence, which is derived from predictability and assurance, trust involves a leap of faith, accepting the risk of potential disappointment as an integral component of the relationship dynamic \cite{Luhmann2000}. The debate between psychologists on trust's essence, whether it is an emotional leap of faith or a rational and goal-oriented choice, underscores its multifaceted nature \cite{Smith2005, Giddens2007, Simmel2011, Taddeo2010}. Trust can be viewed both as a deep-seated psychological attitude and a pragmatic evaluation of the benefits and risks of relying on another \cite{Taddeo2010}.
The process of evaluating trustworthiness is context dependent. Indeed, in personal relationships, trust is built through direct and repeated interactions, enabling the trustor to form a robust perception of the trustee's reliability and intentions \cite{Dasgupta2000, Ellickson1994}. 

Conversely, institutional trust hinges on the perceived legitimacy conferred by formal credentials or societal recognition. In technological contexts, this assessment shifts towards confidence in the system's adherence to predefined rules and reliability \cite{Mitchell2005}. This transition from personal judgment to systemic predictability reflects a broader societal trend: people often perceive technologically driven institutions as more 'trustworthy' than their human-led counterparts, attributing to technology the capacity to build confidence through its predictability and reliability \cite{Lustig2015}. This perception underscores the role of technology in fostering a sense of assurance and reliability that, in turn, serves as a foundation for trust \cite{DeFilippi2020}.
Unlike trust, which is borne out of a decision made under uncertainty and carries inherent vulnerability, confidence is derived from the stability and continuity observed in past experiences or the established credibility of third-party experts \cite{Pavlickova2013, Luhmann1979}. This foundation of predictability significantly reduces perceived risk, delineating confidence from the vulnerabilities associated with trust \cite{Luhmann1979}. The essence of confidence is further elucidated by Simmel's concept of "weak inductive knowledge," which posits that confidence is rooted in broad experiences or the trustworthiness of experts outside one's direct personal knowledge \cite{Simmel2011}. Therefore, confidence doesn't require the individual to make a vulnerable leap of faith or engage in the active decision-making process characteristic of trust \cite{DeFilippi2020}. Instead, it represents an assured cognitive state, shaped by a history of reliability and the expectation that future events will unfold in a manner consistent with past occurrences \cite{Seligman1998}. 

This form of assurance, grounded in the objective assessment of systemic reliability rather than subjective judgment, underscores the fundamental difference between confidence and trust.
The relationship between trust and confidence is characterized by a dynamic interplay, where confidence can act as a platform for the development of trust. The more confidence there is in a higher-order system, the easier it becomes for individuals to establish trust relationships with entities operating within that system. For example, confidence in the efficacy of a healthcare system can enhance patients' willingness to trust individual healthcare providers \cite{DeFilippi2020}. Similarly, confidence in the integrity of a financial system can encourage individuals to engage more readily with financial institutions \cite{Putnam2000}. This nuanced understanding of trust and confidence, highlighting the responsibility inherent in trust decisions, the impact of technology in shaping perceptions of trustworthiness, and the symbiotic relationship between trust and confidence, enriches the discourse on these critical components of social and systemic interaction.\\

This exploration of trust and confidence has profound implications for assessing employees' wellbeing in organisational contexts. Well-being assessments, inherently reliant on employees' willingness to share honest feedback, necessitate an environment where trust in the confidentiality and ethical use of data is paramount. However, the efficiency and reliability of these assessments hinge on confidence in the systems used for gathering and analysing feedback. The challenge lies in balancing the need for transparent, secure systems that protect employee data (confidence) with fostering an organisational culture that values and acts on employee feedback in a trustworthy manner (trust).
In essence, the successful implementation of well-being assessments requires a dual approach: enhancing system reliability to bolster confidence while simultaneously cultivating a culture of trust where employees feel safe to express their genuine concerns and needs. The delicate interplay between trust and confidence in this context underscores the complexity of managing human dynamics in organisational settings \cite{Mitchell2005}, where the ultimate goal is to achieve a harmonious environment that supports both the individual's and the organisation's wellbeing. This complex relationship underscores the necessity of innovative solutions that can bridge these conceptual divides \cite{Lustig2015}. 
While traditional methods have struggled to ensure the confidentiality, integrity, and reliability of well-being assessments, the emergence of blockchain technology offers a promising avenue to address these concerns directly. As we shift from understanding the foundational dynamics of trust and confidence, the following section explores how blockchain technology can serve as a transformative tool in the landscape of well-being assessments, while acknowledging its cultural limitations.

\section{Blockchain technology and well-being assessment}
Blockchain technology is a decentralised, distributed ledger system that has transformed how digital transactions are conducted, verified, and securely recorded. 
From a technological perspective, it seeks to empower anyone with an internet connection to securely transfer valuable digital assets—including currency, software code, documents, or survey responses—while ensuring unmatched security and integrity \cite{Casino2019}. 
Blockchain operates on a decentralised, peer-to-peer network where data and its change history (also considered as a series of transactions) are securely organized in a chain of cryptographically linked blocks. Each block contains a unique hash, i.e. a unique identifier generated by a secure mathematical algorithm for the data that ensures the integrity of the information, making it resilient against both unintentional and malicious manipulation while remaining accessible to all participants on the network \cite{Zheng2017}. 
This methodical process of adding transactions to the blockchain ensures that all transaction records are permanent and tamper-evident, providing a clear, auditable trail of activity within the network. The principle of decentralisation is fundamental to blockchain technology. Unlike traditional centralised systems, where a single entity has control over the transaction ledger, a blockchain environment allows this ledger to be maintained concurrently across numerous nodes, eliminating any single point of failure and ensuring that no one entity can unilaterally alter the transaction record \cite{Gatteschi2018}.
Potential applications for blockchain have now expanded far beyond the sole cryptocurrency domain initiated by Bitcoin in 2008 \cite{Nakamoto2008}, encompassing supply chain management \cite{Kim2018}, protection of digital identity \cite{Zwitter2020}, enhancements in financial services \cite{Treleaven2017}, advancements in clinical research \cite{Charles2019}, protection of intellectual property rights \cite{Wang2019}, securing complaint management systems against harassment \cite{Rahman2022}, tracking employee well-being in suppliers’ factories \cite{SHINE2021}, and addressing the deterioration of working conditions in academia \cite{Sicard2022}, among others.\\

\subsection{Blockchain as a confidence machine for well-being assessments}
The essence of blockchain technology, often celebrated for its potential to function without requiring trusted intermediaries, marks a significant paradigm shift in organisational practices. Andreas Antonopoulos characterizes this shift as moving from relying on interpersonal trust to trusting in the algorithmic integrity of blockchain systems \cite{Antonopoulos2014}. This concept, further defined by Kevin Werbach as "trustless trust" \cite{Werbach2018}, suggests that the security and reliability of transactions, by extension, the assessment processes within organisations, are ensured through deterministic computational means rather than through traditional trust dynamics. This argument, primarily negative, focuses on eliminating the need for trust to facilitate interactions that might otherwise be hindered by skepticism or fear of exploitation \cite{Das2004}.

Blockchain instills confidence through several mechanisms, such as its mathematical foundation (cryptographic hash functions) that obviates the need for traditional forms of trust. This mathematical reliability promises high predictability and security, as evidenced by the robustness of blockchain protocols like Bitcoin, which has remained secure against hacks despite rigorous scrutiny \cite{Seligman1998}. Its consensus mechanisms are also crucial for validating transactions and maintaining the ledger’s integrity without the need for traditional centralised systems. The consensus process, which might involve protocols such as proof of work or proof of stake, ensures that all network participants agree on the ledger’s state, thereby preventing fraud and ensuring that each transaction is accurate and secure.

These aspects can naturally translate in organisational contexts, where the integrity of employees' well-being data is paramount. Since the technology operates independently of any centralised authority, it can be perceived as a less corruptible alternative to traditional mechanism for monitoring and enhancing employees’ wellbeing. In fact, employees often suffer from biases, inaccuracies, and a lack of transparency. Blockchain introduces a paradigm where the assessment of employees' wellbeing can be conducted in a manner that is both immutable and transparent, ensuring a fair and accurate representation of employees' conditions \cite{Snow2014, Benchoufi2017}. In the organisational context, the shift towards blockchain technology for well-being assessments signifies a move towards establishing a robust framework of confidence, one where the integrity, transparency, and immutability of blockchain offer a solid foundation for reliably capturing and reflecting the true state of employee well-being \cite{Lustig2015}.
By enhancing the degree of confidence in the systems used for assessing employees' well-being, blockchain indirectly reduces the reliance on trust, thereby streamlining interactions and assessments by mitigating perceived risks \cite{DeFilippi2020}. 
Through checks, balances, and transparency, akin to the principles advocated by Hume \cite{Hume1987} and Hardin \cite{Hardin2002}, blockchain can foster a more secure environment for these assessments. However, this technological pivot supporting a foundation of confidence does not obviate the need for trust entirely \cite{Maurer2013, Nickel2015}. But to what extent does truth play a crucial role in the effective conduct of well-being assessments?

\subsection{Limitations of the confidence machine – the need for trust}
While blockchain technology introduces a significant enhancement in the security, transparency, and integrity of data management within well-being assessments, it alone is not sufficient to address all the intricacies involved. 
The technology is fundamentally seen as a "confidence machine," adept at creating a secure and immutable ledger where data alterations are transparent and traceable \cite{DeFilippi2020}. This capability is undoubtedly valuable in environments where the accuracy and permanence of data are critical. However, well-being assessments in organisations demand more than just data integrity, they require a deep understanding of the human elements that blockchain technology can hardly provide.

In health and social care, modernisation policies have emphasized efficiency and convenience, which often parallels the confidence that blockchain brings through its technological capabilities \cite{Sevenhuijsen1998}. However effective well-being assessments require more than just efficient data processing; they require trust in the intentions and behaviors of those who not only design the surveys but also interpret and act upon the data \cite{Maurer2013}. 
These assessments often deal with sensitive information about employees' mental and physical health, where the context and subtleties of human experience play crucial roles \cite{Blum1991}. The challenge in well-being assessments, similar to that in health care, lies in ensuring that the system’s efficiency does not undermine the quality of human attention that is critical to meaningful interactions and interventions \cite{Smith2005}. Employees must trust that the administrators of these assessments will handle their data with care and use it to genuinely enhance workplace well-being. 
In organisational contexts, employees may withhold full participation in well-being assessments if they do not trust how their data will be used. Without this trust, even the most accurate data can fail to lead to effective solutions, as employees might not see the assessments as genuinely aimed at improving their workplace well-being but rather as a tool for surveillance or performance evaluation. Therefore, the collection of their true state of wellbeing will be nearly impossible because if this absence of trust \cite{Winner1980}.

Moreover, blockchain's role in enhancing data security does not automatically translate to an increase in trust among employees. Security, transparency and trust are related but distinct concepts; secure and transparent data can still be used in ways that defeat the interests of the data subjects, both in terms of representation of stigmas and addressing power imbalance. For instance, data collected securely via blockchain could still be used to implement changes that are perceived as invasive or punitive, if not tempered by trust in the intentions behind these changes \cite{Josang2007}. Therefore, organisations must not only implement blockchain to leverage its strengths in enhancing the confidence in system capabilities but also actively work to foster interpersonal trust. This involves transparent communication about how data is collected, interpreted, and used, as well as involving employees in the processes that their data informs.\\ 

Overall, it is critical to recognize that the absence of, or imbalance between, trust and confidence is a significant issue in the current landscape of well-being assessments. Increasing confidence through technological surrogate like blockchain can certainly bolster the process, but it does not address the full spectrum of needs in these assessments. Trust in the people behind the technology, the administrators interpreting the data and making decisions based on it for example, is equally important. While blockchain puts confidence into the process, addressing the mechanical aspects of data security and integrity, psychology, the human response to and engagement with these systems, does not stop at the technological process. Well-being assessments are as much about understanding and responding to human needs and nuances as they are about collecting data. Thus, without trust, the most sophisticated systems may still fall short of their goal to genuinely improve well-being in the workplace. This dual need for both trust and confidence underscore the complex nature of implementing effective well-being assessments in modern organisational environments.

\begin{table*}[h]
\begin{center}
\begin{tabular}{| p{5cm} | p{5cm} | p{4cm} |}
  \hline
  \textbf{Relational-Cultural Theory} & \textbf{Co-Production} & \textbf{Alignemnt} \\
  \hline \hline
  \textbf{Relational Authenticity:} Being genuine and transparent in relationships fosters trust, deepens connection, and enhances collaboration. Authenticity allows individuals to bring their full selves into the relationship, enabling honest communication. 
  & 
  \textbf{Authentic Collaboration:} Co-production requires stakeholders to be open about their needs, capacities, and expectations. Transparent and genuine communication helps build trust and ensures that all participants feel heard and respected.
  & 
  Both rely on openness, trust, and transparency for effective collaboration. Authenticity strengthens the relational foundation and promotes meaningful contributions. \\
  \hline
  \textbf{Perceived Mutuality:} Mutuality involves shared responsibility, reciprocal influence, and a sense of interdependence in the relationship. Each person recognizes their impact on the other and works toward a shared outcome.
  &
  \textbf{Shared Power and Decision-Making:} Co-production fosters mutuality by giving all stakeholders equal responsibility and influence in the design, delivery, and evaluation of solutions. Each participant’s perspective is valued, and decisions are made collectively.
  & 
  Both emphasize reciprocal influence and shared responsibility, ensuring that participants work together to shape outcomes. Mutual respect and interdependence are central to this dynamic. \\
  \hline
  \textbf{Relational Connection:} Strong, empathetic connections are essential for fostering growth, emotional well-being, and collaboration. Relational connection brings a sense of belonging and solidarity, encouraging deeper engagement in the relationship.
  &
  \textbf{Building Trust and Relationships:} Co-production thrives on the creation of strong relationships between stakeholders, where trust and a sense of shared purpose are key. These connections encourage long-term engagement and commitment to the process.
  & 
  Both emphasize the importance of building strong, empathetic relationships that foster trust, belonging, and sustained engagement. Connection is essential for collaboration and mutual understanding. \\
  \hline
  \textbf{Relational Empowerment:} Relationships should foster empowerment, where both individuals feel supported and able to express their needs and capabilities. Empowerment is about enabling each person to have agency and voice within the relationship.
  &
  \textbf{Empowering Participants:} Co-production empowers all participants, particularly marginalized groups, by giving them a significant role in shaping solutions. The process promotes agency and ensures that everyone has an equal voice in decision-making.
  & 
  Both focus on empowering individuals through active participation and shared decision-making, allowing people to express their perspectives and have a tangible impact on outcomes.\\
  \hline
\end{tabular}
\caption{Description of the dimensions of the relational-cultural theory and co-production frameworks relevant to well-being assessment and their alignment regarding relational strategies for trust-building}
\end{center}
\label{tab:1}
\end{table*}

\section{Co-production as a cultural element for building trust in the digital realm}
\subsection{Relational cultural theory and the role of co-production in building trust}

As discussed previously, trust building remains a critical element that can be easily overlooked in implementation strategies that aim to influence key implementation outcomes such as acceptability, adoption, fidelity, reach, and sustainability \cite{Proctor2011}. This is particularly true when implementation strategies are based on blockchain technologies. 
To effectively address the issues of trust and confidence, blockchain-based strategies should incorporate two core mechanisms: relational strategies and technical strategies. Although blockchain technology, seen as a 'confidence machine', can address the technical side - defined as strategies that aim to build trust by demonstrating knowledge, reliability and competence to support the goals of the team - it does not address the relational side. 
 Relational strategies can be seen as efforts to build trust by representing stigmas and addressing power differentials among implementation teams and stakeholders, thus strengthening the quality, mutuality, and reciprocity of their interactions \cite{Metz2022}.

To circumvent this limitation and strategically foster trust within implementation teams and stakeholders, Relational Cultural Theory (RCT) provides a valuable theoretical framework, supporting the idea that understanding others' perspectives increases a sense of mutual interdependence and leads to positive emotional responses among individuals in relationships \cite{Leeman2017}. 
RCT primarily focuses on creating and maintaining growth-fostering connections through interpersonal relationships, examining how personal growth and emotional health are shaped by relational dynamics, particularly through empathy-driven exchanges, demonstrations of authenticity, and mutual empowerment. RCT aims to flatten hierarchical structures and challenge power imbalances that affect trust, offering a solid relational framework based on several relational dimensions, including relational authenticity, perceived mutuality, relational connection, and relational empowerment (see Table \ref{tab:1}).\\

In this context, co-production aligns with RCT’s core assumption that meaningful outcomes are achieved through collaborative, interdependent relationships.
Co-production, first coined by Ostrom in the 1970s and defined as “the role of individual choice on decisions influencing the production of public goods and services”, is still a relatively new concept in research and practice \cite{Ostrom1996}. In essence, if executed well, co-production allows for the redressing of power imbalances, providing a foundation for relational ethics and confronting complexities head-on, centralising key principles, such as inclusivity/diversity, respecting knowledge and reciprocity \cite{Filipe2017, Tan2020}.
Research and practice involving co-production are generally centred around three broad premises \cite{SimsSchouten2024}. First, the right to be involved in decisions affecting oneself, second the need to improve the value of a project, and third, the requirement to enhance knowledge on a topic \cite{Turakhia2017}. Co-production can promote justice and lead to new knowledge, thereby fundamentally democratizing the relationships between the different parties: researchers and research participants \cite{SimsSchouten2025}. Central to this is the notion that co-production facilitates equal collaboration between ‘experts by experience’ and ‘experts by qualification’, culminating knowledge and freedom of expression, and revealing positions and positionality \cite{Rikala2020}.
Thus, both RCT and co-production emphasize that productive exchanges occur when two or more people jointly create benefits that cannot be achieved alone \cite{Thye2002}. 
Co-production, like RCT, values mutual engagement and equality, operating on the principle that those with lived experience are uniquely positioned to contribute to designing effective solutions. This partnership-driven approach enhances collective outcomes by building on the same relational dynamics that RCT highlights.

By implementing co-production in the design of well-being assessment in organisational contexts, implementation teams and stakeholders can engage in co-learning and co-design processes, enabling them to negotiate and build trust and respect for all perspectives, including those at risk of being excluded from dialogue due to existing stigmas such as race, ethnicity, language, or status, among others.
Co-production offers an inclusive relational framework based on the dimensions of authentic collaboration, shared power and decision-making, relationship building, and participant empowerment, which closely align with the RCT dimensions of relational authenticity, perceived mutuality, relational connection, and relational empowerment, respectively (see Table \ref{tab:1}). In particular, both frameworks focus on creating growth-fostering, equitable relationships where all participants have an equal role, addressing power imbalances, and building connections that promote trust.\\

\subsection{Co-produced Decentralised wellbeing assessment framework (CoDeWe)}
Integrating the cultural component of trust with the technological component of confidence is essential for establishing a dual cultural-technological framework to design a co-produced decentralised well-being (CoDeWe) survey based on blockchain technology, where employees' well-being can be trustfully assessed.
\begin{figure*}
\begin{center}
\includegraphics[width=18cm]{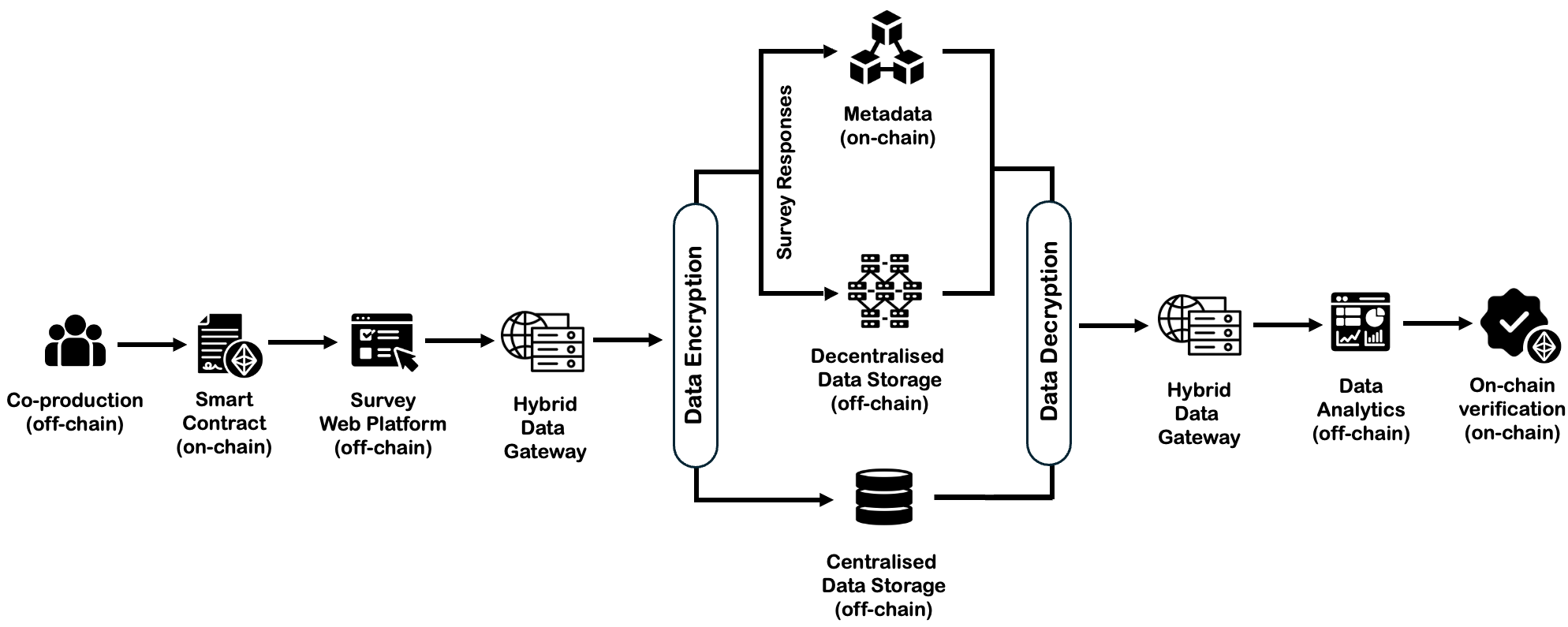}
\end{center}
\caption{\textbf{Overall architecture of the CoDeWe framework showing the connection between the cultural component (co-production) and technological component (on-chain metadata, smart contracts, digital signatures, decentralised data storage) needed to balance cultural trust with technological confidence.} 
The different dimensions of the surveys are co-produced off-chain using capacity building, technical facilitation, participatory design and feedback loops to identify and address potential stigmas and power imbalance among the different stakeholders (step 1). 
The finalised questions and rules of the surveys are translated on-chain into a smart contract deployed on a public blockchain, such as Ethereum (step 2). 
The respondents access the survey off-chain using a web interface (step 3). 
After completing the survey, the respondent digitally signs their responses using their unique cryptographic key to verify their authenticity (step 4).
The actual survey responses are stored off-chain in a decentralised data storage (step 5). 
The unique hash identifier generated for the survey responses along with the respondent's digital signature and a timestamp constitute the metadata that is recorded on-chain on the blockchain. Additionally, a centralised data storage, such as MySQL, provides efficient querying and analysis of responses off-chain (step 6).
The administrator periodically retrieves responses from the off-chain centralised database.  They analyse the data, generating reports and visualisations based on the survey results. After analysis, they generate a summary hash for the analysed data. which is stored on-chain to provide a verification point for users (step 7 and step 8).
}\label{fig:1}
\end{figure*}
As shown in Figure \ref{fig:1}, the overall architecture of the CoDeWe framework can be divided into two parts: a cultural component for co-producing the various dimensions and questions of the survey (Capacity Building, Technical Facilitation, Participatory Design, Feedback Loops), and a technological component for data storage (decentralised database management, e.g. IPFS), data security (Blockchain, e.g. Ethereum), and data queries and analysis (centralised database management system, e.g. MySQL). The workflow of CoDeWe consists of the following steps:

\textbf{Step 1 - Survey Co-production:} The survey is co-designed with input from participants, ensuring their lived experiences are incorporated into the survey structure and questions. This stage involves a series of collaborative workshops or sessions where stakeholders, including the survey administrator, researchers, and participants, jointly contribute to defining the objectives, survey questions, and parameters. Throughout the co-production process, discussions focus on identifying and addressing potential stigmas that may impact participants’ willingness to respond honestly, with strategies developed to frame questions in a way that minimizes stigma and encourages open, honest participation. Key aspects of co-production include capacity building, which provides participants with the necessary knowledge and tools to meaningfully contribute to survey design; technical facilitation, which offers support to both administrators and participants to understand how digital technologies like blockchain, decentralised data storage, and cryptographic signature help secure and verify their responses; participatory design, allowing participants to shape the questions and structure of the survey to reflect their needs while being sensitive to stigmas; and feedback loops, which continuously gather input from participants and stakeholders during the design phase to refine and improve the survey based on their suggestions and concerns. 

\textbf{Step 2 - Survey Setup:} A smart contract is created and deployed on the blockchain network, which defines the rules of the survey and handle the co-production process, storing lived experiences of participants in the co-design sessions, the finalised questions, and the agreed survey parameters. The administrator digitally signs the smart contract (e.g. survey parameters such as questions, rules, etc.) using their unique cryptographic key (i.e. private key) to prove authenticity and  confirm the final structure and design. The signature and survey parameters (including a hash of the parameters) are stored on the blockchain to ensure that the survey cannot be altered after creation.

\textbf{Step 3 - Survey Distribution:} The survey link or instructions are shared with potential respondents. Respondents are made aware of the co-production aspect and that their responses will be recorded in a secure, verifiable manner, and that no personal identifiable information will be recorded.

\textbf{Step 4 - Respondent Submission:} Respondent fills out the survey using a web interface. After completing the survey, the respondent digitally signs their responses using their private key. This signature verifies the authenticity of the responses and confirms that they have not been tampered with. A hash of the responses is generated to create a unique identifier for the submission.

\textbf{Step 5 - Data Storage:} The actual survey responses are stored off-chain in a decentralised data storage such as IPFS. The IPFS network generates a content-addressed hash for the responses, which serves as a unique identifier. The response hash, along with the digital signature, is sent to the blockchain network. The blockchain records on-chain the associated metadata, including the hash of the survey responses, the IPFS hash (link to the actual data), the respondent's digital signature, and a timestamp and other relevant metadata (e.g., response ID, survey ID).

\textbf{Step 6 - Verification and Integrity Check:} The blockchain serves as an immutable ledger of all submissions. It allows anyone to verify whether a specific response was submitted by checking the hash stored on-chain and the authenticity of the response by verifying the respondent's digital signature against their public key (i.e. a cryptographic key created with the corresponding private key that is shared openly and used to verify signatures or encrypt data). If there is a need to check whether any responses were excluded from analysis, the administrator can retrieve the list of all response hashes stored on-chain and compare this list against the hashes of responses that were analysed (stored in the centralised database, e.g. MySQL).

\textbf{Step 7 - Data Analysis:} The administrator periodically retrieves responses from IPFS for analysis and stores them in a MySQL database. This allows for fast querying and analytical capabilities. They analyse the data stored in MySQL, generating reports and visualisations based on the survey results. After analysis, they generate a summary hash (or Merkle root) for the analysed data. They store this summary hash on-chain to provide a verification point for users. Users can then verify if their responses were included in the analysis by checking if their response hash matches any included in the analysed dataset.

\textbf{Step 8 - Reporting and Feedback:} After analysis, results and insights can be shared with respondents or stakeholders. Digital signatures can also be used to sign off on the final report, ensuring that the published results come from a verified source.

\section{Discussion and Conclusion}

While the Principles for Responsible Investment (PRI), endorsed by the United Nations and recognized as a leading advocate for responsible investment, identify mental health and healthcare accessibility as two of the four critical social concerns that emerged following the pandemic, the growing trend of integrating well-being into ESG reporting is likely to reshape corporate approaches to employee well-being assessment. This trend reflects a shift towards acknowledging the multidimensional nature of health, as defined by the World Health organisation, which extends beyond the traditional ESG focus on injury and disease prevention, prompting companies to capture a more holistic view of their employees' experiences.\\

In this context, the increasing investor interest in companies’ well-being programs and the growing expectations of Generation Z are driving demands for more transparent disclosure of not only the existence but also the inclusiveness of these programs. 
Companies are therefore incentivized to innovate by adopting digital solutions and data analytics to enhance transparency, efficiency, and effectiveness in managing ESG-related issues, thereby contributing to sustainable and responsible business practices. However, this digital shift also raises critical issues regarding the role of technology in striking a delicate balance between how trust and confidence are culturally and/or technologically achieved.\\

Building on Relational Cultural Theory and blockchain technology management, we explored how decentralised well-being surveys can be supported by cultural practices that foster and sustain trust, enabling the effective integration of blockchain technology into well-being assessments. We explained how trust-building can be culturally achieved through co-production, which helps to address power imbalances between the implementation team and stakeholders.
Specifically, we presented a dual cultural-technological framework and the associated workflow, providing conceptual clarity on how the technological implementation of confidence can align with the cultural development of trust. This ensures that blockchain-based decentralised well-being surveys are not only secure and reliable, but also perceived as trustworthy tools for improving workplace conditions.\\

From a technical perspective, blockchain technology is still considered relatively new, with its own challenges and drawbacks, including regulatory risks and a lack of well-defined use cases, which are hindering mass adoption. However, it is increasingly having a positive impact across many industries and is gaining recognition in the academic space.
In the context of workplace well-being assessments, it is crucial to address both the challenges and potential solutions for complying with Regulation (EU) 2016/679 of the European Parliament and of the Council of 27 April 2016 on the protection of natural persons with regard to the processing of personal data and the free movement of such data, repealing Directive 95/46/EC (General Data Protection Regulation, GDPR), particularly the right to erasure (‘right to be forgotten’) as outlined in GDPR Article 17 \cite{Finck2018}. The personal information collected in these assessments comes from questionnaires specifically designed to evaluate employee well-being. As the design of these questionnaires is co-produced with stakeholders, the handling of personal data must be integrated into the technical facilitation stage of the co-production process.
While GDPR does not apply to anonymized data that cannot be traced back to an individual, cryptographic hash functions—fundamental to blockchain technologies—only achieve pseudonymization and do not fully comply with GDPR when personally identifiable information (PII) from the questionnaires is stored on-chain \cite{Finck2020}. To protect individual privacy, it is advisable to avoid or limit the use of on-chain storage for PII. Instead, implementing decentralised well-being assessment protocols that do not explicitly store PII, along with using temporary digital signatures, would provide a sustainable and GDPR-compliant solution. This approach ensures that the co-produced design of the surveys is aligned with privacy and data protection regulations from the outset.\\

While the proposed framework and workflow offer a comprehensive approach to integrating blockchain technology with co-produced well-being assessments in compliance with GDPR, it is crucial to test and validate this system in workplace settings. Some existing well-being assessment methods may already integrate smoothly with blockchain technology and co-production, and testing will allow us to identify which methods align most effectively with the framework. Practical implementation will also help assess the system's ability to address key challenges such as privacy protection, trust-building, and data security. Field-testing in diverse workplace environments will provide essential insights, enabling further refinement and optimisation to ensure that decentralised well-being assessments are not only theoretically sound but also operationally efficient, contributing meaningfully to improving workplace conditions.

\section*{Acknowledgements}
FS thanks Maxime Nicolas and Kai Yu for useful discussions.

\bibliographystyle{elsarticle-harv} 
\bibliography{bibliography}

\end{document}